\documentclass[aps,prc,twocolumn,nofootinbib]{revtex4}
\usepackage{graphicx}
\usepackage{amsfonts,amsmath,amssymb,bm}

\def\beq{\begin{equation}}
\def\eeq{\end{equation}}
\def\bea{\begin{eqnarray}}
\def\eea{\end{eqnarray}}
\def\nn{\nonumber}
\def\lan{\langle}
\def\ran{\rangle}

\begin{document}

\title{Dissipative hydrodynamics coupled to chiral fields}

\author{J. Peralta-Ramos\footnote{jperalta@ift.unesp.br} and 
G. Krein\footnote{gkrein@ift.unesp.br}}

\affiliation{Instituto de F\'isica Te\'orica, Universidade Estadual Paulista, 
Rua Doutor Bento Teobaldo Ferraz, 271 - Bloco II, 01140-070 S\~ao Paulo, SP, Brazil}

\begin{abstract}
Using second--order dissipative hydrodynamics coupled self--consistently to the linear 
$\sigma$ model we study the $2+1$ dimensional evolution of the fireball created in 
Au$+$Au relativistic collisions. We analyze the influence of the dynamics of the chiral 
fields on the charged-hadron elliptic flow $v_2$ and on the ratio $v_4/(v_2)^2$ for a 
temperature--independent as well as for a temperature--dependent viscosity--to--entropy 
ratio $\eta/s$ calculated from the linearized Boltzmann equation in the relaxation time approximation. 
We find that $v_2$ is not very sensitive to the coupling of chiral sources to the hydrodynamic evolution, but the temperature dependence of $\eta/s$ plays a much bigger
role on this observable. On the other hand, the ratio $v_4/(v_2)^2$ turns out to be much more 
sensitive than $v_2$ to both the coupling of the chiral sources and the temperature dependence
of $\eta/s$.  
\end{abstract}

\maketitle

\section{Introduction}

One of the most surprising outcomes of the experiments conducted at the Relativistic
Heavy Ion Collider (RHIC) is the discovery that high--energy collisions of heavy
ions produce strongly interacting hadronic matter (quark-gluon plasma -- QGP) that 
evolves as a low--viscosity fluid~\cite{{RHIC-disc1},{RHIC-disc2},{RHIC-disc3},{RHIC-disc4},
{RHIC-disc5},{RHIC-disc6}}. This came as a surprise because early expectations, motivated 
mainly by the asymptotic freedom property of Quantum Chromodynamics (QCD), were that RHIC 
would create a gas--like system of weakly interacting quarks and gluons. 

The strongly interacting nature of the QGP is revealed through the description of measured momentum
anisotropies via relativistic viscous hydrodynamics calculations. The momentum anisotropies 
are encoded in the Fourier moments $v_2$, $v_3$, $v_4$, $\cdots$ of the measured azimuthal
distribution of particles, and are interpreted as being the translation to momentum space 
of the initial spatial eccentricity of non central collisions~\cite{olli}. 
A~weakly--interacting, gas--like fluid would have no mechanism to induce such a translation 
into momentum anisotropies, but they can be produced if the particles of the fluid are 
strongly interacting. 

In any attempt of describing experimental results of a heavy ion collision via a
viscous hydrodynamics model, crucial physics input reflecting the properties of 
the flowing hadronic matter must be supplemented, like the equation of state (EOS) and 
transport coefficients, as shear ($\eta$) and bulk ($\zeta$) viscosities. Although it is 
expected that for the strongly--coupled QGP $\eta$ and $\zeta$ will depend strongly on 
temperature -- see below -- usually in hydrodynamic simulations temperature--independent values 
for these coefficients are assumed throughout the entire evolution. 
The impact of the temperature dependence of transport coefficients 
on momentum anisotropies that are obtained from hydrodynamic models has been recently 
investigated in Refs.~\cite{Nagle:2011uz,niemi,shen,bozek,koide,bulk1,songheinz}. 

These transport coefficients are, in principle, derivable from first--principles QCD calculations at finite
temperature, but in practice, the calculations are extremely difficult at strong coupling.
There is a long history of computations of the EOS via large--scale numerical simulations of
QCD on a lattice and modern calculations are achieving the accuracy needed for reliable use in
heavy-ion physics -- for a recent overview on the subject, see Ref.~\cite{Kanaya:2010qj}. 
On the other hand, lattice QCD calculations of transport coefficients are still in their 
infancy and only recently have results for shear and bulk viscosities been 
obtained~\cite{{visc-latt1},{visc-latt2},{visc-latt3}}. Analytically, calculations have 
been performed within QCD for the shear and bulk viscosities~\cite{{trans-pert1},{trans-pert2}} 
at extremely high temperatures, where perturbation theory in the coupling constant can be 
used on account of asymptotic freedom. Reliable calculations for these quantities can 
also be done at very low temperatures, where matter consists essentially of a very 
dilute gas of pions. There is plenty of experimental information on elementary 
low-energy hadron-hadron interactions available that can be used to constrain detailed 
calculations that go beyond lowest-order in chiral perturbation theory~\cite{{trans-low1},
{trans-low2}}. On the other hand, in the region of intermediate temperatures, 
$ T \sim \Lambda_{QCD}$, where one expects the deconfinement and chiral transitions to 
occur~\cite{Wilczek:1999ym}, perturbative expansions (in the coupling constant or chiral) 
are not applicable. There is lack of experimental information on the appropriate degrees 
of freedom to use and phenomenological approaches are an alternative, like the use of 
relativistic kinetic theory~\cite{{trans-kinetic1},{trans-kinetic2},{trans-kinetic3},{trans-kinetic4},
{trans-kinetic5},kap,jorgeres}. 

Assuming the validity of a relativistic viscous hydrodynamical description, experimental 
extraction of transport coefficients should be possible in principle from comparison 
with measured momentum anisotropy patterns. Moreover, as suggested in Ref.~\cite{CKMcL},
such a comparison with experimental data would allow to pinpoint the location of the QCD 
phase transition or of a crossover from hadronic to quark-gluon matter. In this context, 
in addition to the hydrodynamic degrees of freedom related to energy-momentum conservation,
degrees of freedom associated with order parameters of broken continuous symmetries must 
be considered as well since they are all coupled to each other. Particularly important to 
the quest of determining possible signals of the chiral transition is the coupling of 
chiral fields to the fluid-dynamical modes. The present work is a first step towards the 
study of effects of such a coupling on momentum anisotropies within a viscous hydrodynamic
description.

Specifically, in the present paper we study, in the context of second--order dissipative
relativistic hydrodynamics, the influence of the long--wavelength modes of chiral fields 
on the expansion of the fireball created in Au$+$Au collisions at $\sqrt{s_{NN}} = 200$~GeV. 
In particular, we aim at studying the effect of the coupled evolution of chiral degrees of 
freedom on the flow asymmetry characterized by $v_2$ and $v_4/(v_2)^2$, when finite viscosity
is taken into account within a simple microscopic model for the chiral condensate. 

In our 
model the flowing matter consists of a fluid of quarks and antiquarks in local thermal
equilibrium that evolves according to a second--order dissipative hydrodynamic
model~\cite{{hyd1},{hyd2},{hyd3}} with temperature-dependent speed of sound and transport
coefficients. 
Starting from a high temperature state with
approximately restored chiral symmetry, the system is evolved towards a state where the 
symmetry is spontaneously broken. The quark-antiquark fluid interacts locally with the 
chiral fields of the linear $\sigma$ model (LSM)~\cite{LSM}. We compare the results 
obtained for hadronic observables when the chiral fields are included or not as sources 
in the hydrodynamic equations. 

Similar models for the dynamics of quarks coupled to chiral fields were considered before. 
Mishustin and Scavenius~\cite{mish} constructed a relativistic mean field {\it ideal} fluid 
dynamical model based on the LSM, while Abada and Birse \cite{ab} used the Vlasov equation to
describe the quarks coupled to the LSM. Both studies focused on the evolution of the chiral
fields rather than on the influence of their dynamics on hadronic observables, which is the 
main focus here. Son~\cite{sonchiral} and Pujol and Davesne~\cite{pujol} have developed 
modified hydrodynamic theories including additional (chiral) symmetry-breaking 
hydrodynamic variables, which were subsequently applied by Lallouet et al.~\cite{lal} to 
the study of Bjorken flow in heavy--ion collisions. Later on, Paech et al.~\cite{paech1,paech2} 
coupled 3+1 {\it ideal} hydrodynamics to the LSM including fluctuations of the chiral fields 
and studied the behavior of the azimuthal momentum asymmetry at nonzero impact parameter. 
More recently, Nahrgang and Bleicher~\cite{Nahrgang:2010fz} included phenomenological
dissipation and noise terms in the equation of motion of the $\sigma$ field of the LSM 
coupled to ideal hydrodynamics. They focused on sigma fluctuations and found that at the 
first order phase transition they lead to an increase in the intensity of sigma excitations.
In a subsequent work, Nahrgang, Leupold, Herold and Bleicher~\cite{Nahrgang:2011mg} 
employed the two-particle irreducible (2PI) effective action formalism to set up an
approximation scheme to derive explicit formulas for the dissipation kernel and the 
noise correlation function within the LSM. While the approximations employed deserve 
careful scrutiny, the formulas derived within the mentioned approximation scheme show
interesting  features of dissipation and noise on the relaxation dynamics of the chiral 
dynamics and point to the importance of including reheating of the quark fluid by energy
dissipation from the chiral fields, as shown in the very recent papers by Nahrgang, Leupold 
and Bleicher~\cite{Nahrgang:2011mv} and Nahrgang, Bleicher, Leupold, and 
Mishustin~\cite{Nahrgang:2011vn}. The work of Plumari et al.~\cite{Plumari:2010ah} is more 
in the spirit of ours in the present paper, in that they have investigated the role of the
chiral transition on $v_2$. Specifically, the authors have solved numerically the 
Boltzmann--Vlasov transport equations including both two-body collisions and the chiral 
phase transition as given by the Nambu--Jona-Lasinio model. For conditions prevailing in 
Au + Au collisions at $\sqrt{s_{NN}} = 200$~GeV, the authors find a sizable suppression 
of $v_2$ due to the attractive nature of the chiral field dynamics. They also found that, 
if $\eta/s$ is kept fixed, $v_2$ does not depend on the details of the collisional and/or 
field dynamics and in particular it is not affected significantly by the chiral phase
transition.

At this point it is appropriate to state the limited scope of our work. First of all,
while we use a viscous hydro model coupled to a workable chiral model, we do not intend 
to extract values of transport coefficients by comparing our results to $v_2$ data.
Actually, extraction of transport coefficients by matching the output 
of hydrodynamic and/or kinetic simulations to data is not at all straightforward, as discussed 
e.g. in Refs.~\cite{luzum,{Heinz:2009xj},{REVS},{Nagle:2011uz}}. In particular, to extract 
$\eta/s$ from $v_2$ measurements, a possible route is to first constrain initial $T_i$ 
and final $T_f$ temperatures  (that are used to run the coupled set of hydro-chiral 
equations) by matching calculated spectra to data and, afterwards, extract $\eta/s$ from 
$v_2$ data -- see e.g. Refs.~\cite{luzum,PRC}. In addition, although we use a semi-realistic
initial condition (we use a Glauber smooth initial condition), a fluctuating, lumpy initial
condition leads to lower elliptic flow along with other interesting 
effects~\cite{lump1,lump2,lump3} that certainly have an impact on the extraction of 
transport coefficients. 

The paper is organized as follows. In Section \ref{coupled} we describe the coupling of
the chiral fields to the $2+1$ hydrodynamic model. In Section \ref{solution} we describe the 
initial conditions and the freeze--out prescription used in the simulations. 
We then show and discuss our results in 
Section~\ref{res}, and conclude in Section~\ref{conc}. In Appendix \ref{lsms} we 
briefly overview the LSM and the calculation of the speed of sound and of the shear 
viscosity from the linearized Boltzmann equation in the relaxation time approximation. 
In Appendix~\ref{cutoff} we discuss the dependence of our results on the cut--off 
that we must impose on the value of $\eta/s$ due to the breakdown of viscous hydrodynamics 
when nonequilibrium effects start to dominate.

\section{Coupled chiral--hydro dynamics}
\label{coupled}

In this section we describe the model of coupling chiral fields to hydrodynamical
variables. The evolution of the chiral fields is described by the LSM ~\cite{LSM} 
and hydrodynamics is described by dissipative hydrodynamic 
equations~~\cite{{hyd1},{hyd2},{hyd3}}. This hydrodynamic formalism goes beyond the 
well--known Israel--Stewart theory~\cite{{isrs1},{isrs2},{isrs3}} in that it includes 
all second--order terms in velocity gradients that can appear in the stress--energy 
tensor of a conformal fluid -- see e.g. Ref.~\cite{betz}. This hydro model has been 
applied to model the expansion of the QGP in Refs.~\cite{luzum} and \cite{PRC}. 

In order to set notation and make the paper self-contained, we start with the
LSM model. The Lagrangian density of the model, given in terms of quark $q = (u,d)$ 
and chiral $\phi_a = (\sigma, \vec \pi)$ fields, is written as (we use the signature
for the metric $g_{\mu \nu} = (+,-,-,-)$ )
\beq
\mathcal{L} = \mathcal{L}_q + \mathcal{L}_\phi ,
\label{lag}
\eeq
with 
\beq
\mathcal{L}_q = \bar{q} \left[ i\gamma^\mu \partial_\mu - 
g \left(\sigma+\gamma_5 \vec{\tau}\cdot \vec{\pi}\right)\right] q ,
\label{L_q}
\eeq
and
\beq
\mathcal{L}_\phi = \frac{1}{2} \left(\partial_\mu \sigma \partial^\mu \sigma 
+ \partial_\mu \vec{\pi} \cdot \partial^\mu \vec{\pi}\right) 
- U(\sigma,\vec{\pi}) .
\label{L_phi}
\eeq
where 
\bea 
U(\sigma,\vec{\pi}) &\equiv& U(\phi) = \frac{\lambda^2}{4} \phi^2 - h_q \sigma - U_0 \nn \\
&=& \frac{\lambda^2}{4}\left(\sigma^2+\pi^2-v^2\right)^2 
- h_q \sigma - U_0 ~,
\eea
is the potential which exhibits chiral symmetry breaking -- with $\phi^2 = \sigma^2 + \pi^2$. 
At the level of approximation we work in the present paper, parameters are fixed as follows.
The vacuum expectation values of the condensates are taken $\lan \sigma \ran = f_\pi$ and 
$\lan \vec{\pi} \ran=0$, with the pion decay constant $f_\pi=93$~MeV. Also, the partially 
conserved axial--vector current relation yields $h_q = f_\pi m_\pi^2$ with $m_\pi=138$ MeV. 
This leads to $v^2= f_\pi^2-m_\pi^2/\lambda^2$. A mass $m_\sigma = \sqrt{2\lambda^2 f_\pi^2 
+ m_\pi^2} \sim 600$~MeV is obtained if $\lambda^2 = 20$. The constant $U_0$ is chosen such 
that the potential energy vanishes in the ground state.  

Following Paech et al.~\cite{paech1,paech2}, we split the energy-momentum tensor as
\beq
T^{\mu\nu} = T^{\mu\nu}_q + T^{\mu\nu}_\phi .
\label{split-T}
\eeq
As mentioned previously, the quarks and antiquarks are assumed to constitute a heat 
bath in local thermal equilibrium and their dynamical evolution will be determined by 
viscous relativistic hydrodynamics, so that:
\beq
T^{\mu\nu}_q = (\epsilon + p) u^\mu u^\nu - p g^{\mu\nu} + \Pi^{\mu\nu} , 
\eeq
where $u_\mu$ is the four-velocity normalized as $u_\mu u^\mu = 1$, $\Pi^{\mu \nu}$ 
is the viscous shear-tensor, and $\epsilon = \epsilon(\phi,T)$ and $p=p(\phi,T)$ are the 
energy density and pressure (by definition) at equilibrium at local temperature 
$T$ -- their explicit expressions are given in Appendix~\ref{lsms}. The contribution 
from the chiral fields to the energy-momentum tensor is given by
\beq
T^{\mu\nu}_\phi = \sum^4_{a=0} \frac{\partial \lan \mathcal{L}_\phi \ran}
{\partial(\partial_\mu \phi_a)}\partial^\nu \phi_a -g^{\mu\nu} 
\lan \mathcal{L}_\phi \ran ,
\label{T_phi}
\eeq
where $\langle \cdots \rangle$ means local thermal average.

The hydrodynamic equations of motion are obtained from conservation of the 
energy-momentum tensor, $D_\mu T^{\mu\nu}=0$, where $D_\mu$ is the geometric 
covariant derivative. We employ Milne coordinates (appropriate for a $(2+1)$ 
flow dynamics) defined by proper time $\tau=\sqrt{t^2-z^2}$ and rapidity 
$\psi=\textrm{arctanh}(z/t)$. The hydrodynamic velocity is $u = (u^\tau,u^x,u^y,0)$ 
and the metric tensor reads $g_{\mu\nu} = (-1,1,1,\tau^2)$. In the transverse plane we 
use Cartesian coordinates $x^i=(x,y)$, so that the only non-vanishing Christoffel 
symbols are $\Gamma^\tau_{\psi\psi}=\tau$ and  $\Gamma^\psi_{\tau\psi} = 1/\tau$. 
Since we assume boost--invariance all quantities are independent of rapidity. The 
conservation equations then read 
\bea 
D\epsilon &=& - \left(\epsilon + p\right) \nabla_\mu u^\mu 
+ \Pi^{\mu\nu}\sigma_{\mu\nu} \nn \\
&& + \, g \left(\rho_s D \sigma + \vec{\rho}_{ps}\cdot D\vec{\pi}\right) ,
\label{eq-e} \\[0.2true cm]
\left(\epsilon + p \right) Du^i &=& c^2_s \left(g^{ij}\partial_j \epsilon 
- u^i u^\alpha \partial_\alpha \epsilon \right) 
- \Delta^i_\alpha D_\beta \Pi^{\alpha\beta} \nn \\
&& + \, g \left(\rho_s \nabla^i \sigma + \vec{\rho}_{ps}\cdot \nabla^i\vec{\pi} \right) ~, 
\label{eq-u}
\eea
where $\Delta^{\mu\nu} = g^{\mu\nu} - u^\mu u^\nu$ is a projector (orthogonal to 
the fluid velocity), $c_s$ is the speed of sound, and $\rho_s = \langle \bar q q \rangle$ 
and $\vec \rho_{ps} = \langle \bar q \gamma_5 \vec \tau q \rangle$ are the local 
thermal averages of the scalar and pseudo-vector chiral densities -- their explicit 
expressions are given in Appendix~\ref{lsms}. It is seen that the chiral densities 
act as sources for the evolution of the hydrodynamic variables $\epsilon$ and $u^\mu$. 
Here, $D=u_\mu D^\mu$ is the comoving time derivative, $\nabla_\mu = \Delta_{\mu\alpha} 
D^\alpha$ is the spatial gradient, $\sigma_{\mu\nu}$ is the shear tensor: 
\begin{equation}
\sigma^{\mu\nu}=\nabla^{\lan\mu}u^{\nu \ran} ~.
\end{equation}
The angular braces around Lorentz indices denotes the spatial, symmetric and 
traceless part of a tensor, i.e. if $A^{\mu\nu}$ is a tensor then 
$A^{\lan\mu\nu \ran}$ means:
\begin{equation}
A^{\lan\mu\nu \ran} = \frac{1}{2} \left(\Delta^{\mu\alpha}\Delta^{\gamma\nu}
+ \Delta^{\mu\gamma}\Delta^{\alpha\nu} 
- \frac{2}{3}\Delta^{\mu\nu}\Delta^{\alpha\gamma}\right) A_{\alpha\gamma} ~.
\end{equation}
The evolution of the shear tensor depends only indirectly on the chiral fields 
and it is given by 
\bea
\partial_\tau \Pi^{i\alpha} &=& -\frac{4}{3u^\tau}\Pi^{i\alpha}\nabla_\mu u^\mu 
- \frac{1}{\tau_\pi u^\tau}\Pi^{i\alpha} + \frac{\eta}{\tau_\pi u^\tau} \sigma^{i\alpha} \nn \\
&& - \frac{\lambda_1}{2\tau_\pi \eta^2 u^\tau}\Pi^{<i}_\mu \Pi^{\alpha> \mu} 
- \frac{u^i\Pi^\alpha_\mu + u^\alpha \Pi^i_\mu}{u^\tau}Du^\mu \nn \\
&& -\frac{u^j}{u^\tau}\partial_j \Pi^{i\alpha} ~,
\label{dpi}
\eea
where $\eta$ is the shear viscosity and $(\tau_\pi,\lambda_1)$ are second-order 
transport coefficients. We note that Israel--Stewart formalism is recovered from these 
equations when $\lambda_1 = 0$.

In the Milne coordinates, the classical equations of motion for the chiral fields 
evolving in the background of the quark fluid read
\bea
\left( \partial^2_\tau + \frac{1}{\tau}\partial_\tau 
- \partial^2_i \right) \sigma 
+ \frac{\delta U}{\delta \sigma} &=& -g\rho_s ~, 
\label{eq-sig}\\ [0.2true cm]
\left( \partial^2_\tau + \frac{1}{\tau}\partial_\tau 
- \partial^2_i \right)  \vec{\pi}
+\frac{\delta U}{\delta \vec{\pi}} &=& -g\vec{\rho}_{ps} ~,
\label{eq-pi}
\eea
where $\rho_s$ and $\vec{\rho}_{ps}$ are the scalar and pseudo-scalar chiral
condensates defined above. 

In this way, the evolution of the chiral fields affects the evolution of the 
quark fluid through the sources in the energy--momentum conservation equations; 
in turn, the quark fluid affects the evolution of the chiral fields through the 
densities ${\rho}_s$ and $\vec\rho_{ps}$.  

\section{Solving the equations}
\label{solution}

We solve the hydrodynamic equations given in Eqs.~(\ref{eq-e}) and (\ref{eq-u}) using the 
publicly available code of Luzum and Romatschke~\cite{luzum}. We need to supply the 
locally temperature-dependent transport coefficients $\eta$, $\tau_\pi$ and $\lambda_1$,  
and speed of sound $c_s$. In addition, we need to specify initial conditions for the energy
density $\epsilon$, velocities $u_1$ and $u_2$, and the components of the shear tensor. 
To solve the equations of motion for the chiral fields, Eqs.~(\ref{eq-sig}) and (\ref{eq-pi}), we 
use a simple finite-difference scheme both in the proper-time and space variables. Here
also, one needs to provide initial conditions for the fields and their derivatives.

In all calculations performed in the present paper, we use a $13~\rm{fm} \times 13$~fm 
transverse plane. In the next subsections we discuss the values used for the input
constants and initial conditions. 

\subsection{Transport coefficients and the speed of sound}

As a prototypical example of a relativistic fluid, we will use $\tau_\pi = 2(2-\ln 2) \eta/{sT}$ 
and $\lambda_1 = \eta / {2\pi T}$, that correspond -- in a gradient expansion at 
second--order --  to a supersymmetric Super--Yang--Mills 
plasma~\cite{hyd1,hyd2,hyd3}. However, as mentioned in the Introduction, we will not use 
$c^2_s =1/3$ and $\eta/s = 1/4\pi$ appropriate to the Super--Yang-Mills plasma. 
The temperature dependence of the speed of 
sound is obtained within the LSM and that of  $\eta$ is obtained using the linearized 
Boltzmann equation in the relaxation time approximation within the chiral model of 
Ref.~\cite{trans-kinetic5}. We discuss these in the Appendix~\ref{lsms}.

We note that the value that we use for the relaxation time 
$\tau_\pi = 2(2-\ln 2) \eta/{sT}$ is not the exact value of the Super--Yang--Mills 
plasma~\cite{jorge1,jorge2} . Rather, it is the value obtained through a second--order gradient expansion, that is equivalent to 
a Taylor expansion of the retarded Green's function that linearly relates a dissipative current and a thermodynamical force. 
The expression for $\tau_\pi$ derived in this way is in general different from the one obtained directly from the first pole of the retarded Green's function.
Moreover, as shown in Refs.~\cite{jorge1,jorge2}, the true dynamics of the long--wavelength and low frequency modes of 
strongly coupled theories is not generally described by relaxation--type equations, i.e., the evolution equation for the shear tensor 
includes terms with second time derivatives of the thermodynamic forces -- see also Ref.~\cite{koidenn}. 

In this connection, it is important to remark that it was shown in previous hydrodynamic simulations (without chiral
fields)~\cite{luzum,PRC} that, for values of $\eta/s\lesssim 0.2$, the results for the hadronic 
observables do not depend strongly on the choices for the second--order transport coefficients, 
provided that $\lambda_1 \neq 0$. The influence of terms with second time derivatives of the thermodynamic forces in the 
evolution of the shear tensor of the QGP is still an open issue that surely deserves further investigation. In this work we will neglect  
such terms and use instead the relaxation--type equation given in Eq.~(\ref{dpi}), that is obtained from a second--order gradient expansion. 

\subsection{Initial conditions}

For the initial transverse velocity and shear tensor we use $u^x=u^y=0$, which implies 
vanishing initial vorticity, and $(\Pi^{xx},\Pi^{xy},\Pi^{yy})=0$. The initialization time 
is set to $\tau_0 = 1$ fm/c. It has been shown before that the 
evolution of the shear tensor $\Pi^{\mu\nu}$ is quite insensitive to the initialization 
values, the difference being appreciable only at very early times (see e.g. Ref.~\cite{REVS}). 
We have verified that the elliptic flow shows very little sensitivity to the initialization 
of the shear tensor as well. The initial temperature at the center of the fireball is 
set to $T_i = 333$~MeV. This value for $T_i$ has been used in previous hydrodynamic 
simulations and leads to kaon multiplicity and $\left\langle p_T \right\rangle$ that are
in good agreement with RHIC data~\cite{luzum,PRC}.

The initial energy density profile is calculated using Glauber's model, in which for a 
given impact parameter $b$ we have 
\begin{equation}
\epsilon(\tau_0,x,y,b) = C \,\sigma \, T_A(x+b/2,y) \, T_A(x-b/2,y) ~,
\end{equation}
where $C$ is a constant chosen such that $\epsilon(\tau_0,0,0,0)$ 
corresponds to a given initialization temperature $T_0$ via the EOS, the cross-section 
$\sigma$ is taken to be $\sigma=40$~mb, and $T_A$ is the 
nuclear thickness function given by
\begin{equation}
T_A(x,y) = \int_{-\infty}^\infty \delta_A (x,y,z) ~dz ~,
\end{equation}
with $\delta_A$ is the density distribution for the gold nucleus, taken to be 
of a Woods-Saxon form 
\begin{equation}
\delta_A(x,y,z) = \frac{\delta_0}{1+\textrm{exp}[(|{\vec x}|-R_0)/\chi]} ~,
\end{equation}
where ${\vec x}=(x,y,z)$, $R_0=6.4$~fm and $\chi=0.54$~fm. The parameter $\delta_0$ is 
chosen such that $\int d^3{\mathbf x}~ \delta_A({\mathbf x})=197$, as appropriate for 
Au nuclei. 

As a reasonable ansatz for the initial condition of the chiral fields we use  
\beq
\vec{\pi}(\tau_0,\vec{r})=0 ~~\textrm{and} ~~
\sigma(\tau_0,\vec{r}) = f_\pi \left[ 1 - e^{ - (r/r_0)^2 } \right] ~,
\eeq
with $r^2 = x^2+y^2$ and $r_0 = 9$~fm. In this way, the chiral condensate nearly vanishes 
at the center where the temperature of the fluid is larger and interpolates to $f_\pi$ where 
the temperature is lower.

\subsection{Freeze-out scheme}
In our simulations we use the isothermal Cooper-Frye freeze--out prescription \cite{cooper} 
in which the conversion from hydrodynamic to particle degrees of freedom takes place in a 
three-dimensional hypersurface. The spectrum for a single on--shell particle with momentum 
$p^\mu=(E,\vec{p})$ and degeneracy $d$ is 
\begin{equation}
E\frac{dN}{d^3 p}=\frac{d}{(2\pi)^3}\int p_\mu ~d\Sigma^\mu~ f (x^\mu,p^\mu) ~,
\label{spec}
\end{equation}
where $d\Sigma^\mu$ is the normal vector on the hypersurface.  The non-equilibrium 
distribution function $f$ is given by Grad's quadratic ansatz~\cite{kinbooks}:
\bea
f(x^\mu,p^\mu)&=& f_0(x^\mu,p^\mu) \nn \\
&& + \, f_0(x^\mu,p^\mu)\left[1\mp f_0(x^\mu,p^\mu)\right]
\frac{p_\mu p_\nu \Pi^{\mu\nu}}{2T^2(p+\epsilon)} \nn \\[0.2true cm]
&\simeq & \left[1+\frac{p_\mu p_\nu \Pi^{\mu\nu}}{2T^2(p+\epsilon)} \right] 
\, \exp \left(\frac{-p_\mu u^\mu}{T}\right) ~, 
\label{fneq}
\eea
where $f_0$ is the Fermi-Dirac distribution. The approximation in the third line holds 
when $p>>T$, and it is used in our simulations. The systematic error of this approximation 
is very small at low tranverse momentum $p_T \lesssim$ 2.5 GeV, so we do not expect our 
results to have a significant error coming from this approximation (see Ref.~\cite{luzum}). 

We calculate the spectra for particle resonances with masses up to 2 GeV and then determine 
the spectra of stable particles including feed-down contributions. For this last step we use 
the AZHYDRO package \cite{AZ1,AZ2,AZ3}. In the present paper we will focus on the elliptic flow coefficient
$v_2$ and also on $v_4$ at central rapidity, which are related to the particle spectra
(including feed-down contributions) by
\begin{equation}
E \frac{d N}{d^3\vec{p}} = v_0(p_T,b)[1+\sum_{n=1}^\infty 2 v_n(p_T,b)\cos(n\varphi)] ~,
\label{v0v2}
\end{equation}
with $\varphi=\textrm{arctan}(p_y/p_x)$ and $p_T=(p_x^2+p_y^2)^{1/2}$. For the freeze--out 
temperature we take $T_f = 130$ MeV. This value for $T_f$ is slightly smaller than the one 
usually employed in viscous hydrodynamic simulations of Au$+$Au collisions of $T_f\sim 140$ 
MeV, but allows us to study a broader range of temperatures near our value of $T_c$ in order to 
determine the influence of chiral fields dynamics on hadronic observables.

\section{Results}
\label{res}

We now go over to discuss our results. We start with a very important input to the hydro
equations, the square of the speed of sound $c_s^2=d p/d \epsilon$. 
In Fig.~\ref{cs} we show $c^2_s$ as predicted by the LSM -- its explicit expression is
given in Appendix~\ref{lsms} -- for three values of the coupling constant, namely 
$g=3.0, 3.2, 3.4$, which correspond to a smooth crossover. The reason to consider these 
values for $g$ is that recent works have shown that the crossover phase transition without 
a thermodynamic region where the sound velocity drops 
to zero leads to a faster time development of the system and helps to reproduce RHIC data 
with hydrodynamic simulations~\cite{cspaper1,cspaper2,prat}. Moreover, Lattice QCD calculations 
also favor a smooth crossover with a relatively soft dip in $c_s^2$ over a first--order 
phase transition (see e.g. Refs.~\cite{tc1,tc2,tc3,tc4,tc5}). In our case, the first--order 
transition is obtained when $g=3.8$. As it can be seen from Fig. \ref{cs}, lowering the value 
of $g$ leads to a softening of the 
crossover and a raise in the critical temperature $T_c$. For large temperatures the conformal 
limit $c_s^2=1/3$ is reached, while for low values $c_s^2$ goes to zero. For $g\sim 3.2$ the 
behavior of $c_s^2$ with temperature is in qualitative agreement with that obtained in Lattice 
QCD calculations \cite{tc1,tc2,tc3,tc4,tc5} and with that favored by hydrodynamic simulations based on different 
interpolation schemes used to join Lattice QCD and hadron resonance gas equations of state (see 
e.g. Refs.~\cite{cspaper1,cspaper2,prat}). For $g>3.4$ or $g<3$, the dip in $c_s^2$ at $T_c$ becomes very sharp 
or fades away, respectively, resulting in a temperature dependence that is in disagreement with 
the one obtained in Lattice QCD calculations.

The critical temperature for these values of the LSM parameters is $T_c \sim 150$ MeV, which is 
smaller than the value $\sim 170$ MeV obtained from Lattice QCD simulations \cite{tc1,tc2,tc3,tc4,tc5}. 
Within the LSM it is not possible to obtain much larger values for $T_c$ without going to very low 
values of $g$. These lower values of $g$ correspond to a very soft crossover with an almost 
inappreciable drop in $c_s^2$ that is not in agreement with the results the Lattice QCD. The 
relatively low value for $T_c$ used here does not constitute a serious limitation to our aim of 
obtaining a qualitative understanding of the impact of the evolution of the chiral fields on 
hadronic observables, since we expect the trends obtained in this paper to hold in more 
realistic scenarios. 

\begin{center}
\begin{figure}[htb]
\scalebox{0.675}{\includegraphics{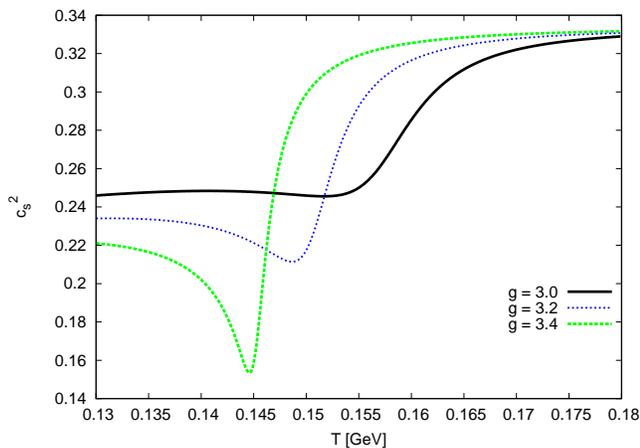}}
\caption{(Color online) Square of the sound speed $c_s^2$ of the LSM as a function of 
temperature, for three values of the chiral coupling: $g=3,3.2,3.4$.}
\label{cs}
\end{figure}
\end{center}

Another input to the hydrodynamic simulation is the ratio $\eta/s$ as a function of temperature, which 
we calculate in the LSM from the linearized Boltzmann equation in the relaxation time 
approximation~(details are given in Appendix~\ref{lsms}). Fig.~\ref{etos} shows $\eta/s$  as a 
function of temperature in a region close to $T_c$, for $g=3,3.2,3.4$. It is seen 
that $\eta/s$ decreases significantly when approaching $T_c$ from below, and remains 
almost constant and rather small at $T>T_c$. This 
behavior is qualitatively similar to that found in other approaches such as the Boltzmann--
hydrodynamics hybrid approach \cite{song}, the NJL model \cite{trans-kinetic5} and Chiral Perturbation 
Theory \cite{nicola}, although, as already indicated, the value of $T_c$ obtained here is 
smaller. It is also seen that when the value of $g$ increases the variation of $\eta/s$ with 
temperature is more abrupt, decreasing faster at $T<T_c$ and thus  remaining small for a larger 
range of temperatures. 

\begin{figure}[htb]
\scalebox{0.675}{\includegraphics{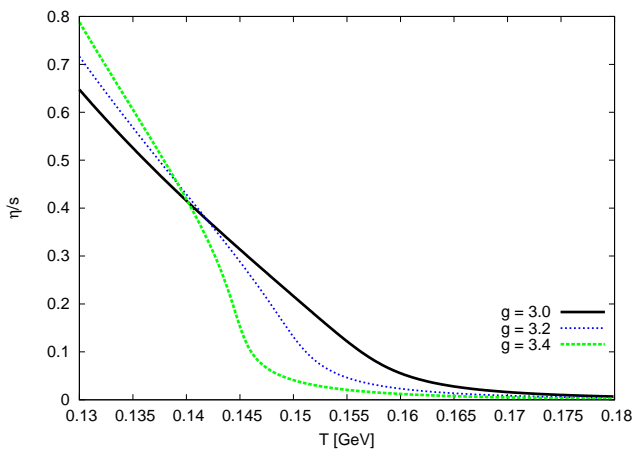}}
\caption{(Color online) $\eta/s$ of the LSM as a function of temperature, for three values 
of the chiral coupling: $g=3,3.2,3.4$.}
\label{etos}
\end{figure}

At this stage it is appropriate to note an important point concerning the rise of $\eta/s$ with 
decreasing temperature in the low temperature stage. By construction, any hydrodynamic description of 
matter necessarily breaks down when the viscosity is large enough that nonequilibrium effects 
are as important, or even more important than the equilibrium ones. For example in the context of Grad's 
quadratic ansatz that is usually employed at freeze--out, this occurs when the nonequilibrium 
correction to the distribution function is comparable -- or exceeds -- the equilibrium one. In the 
context of relativistic heavy--ion collisions numerous hydrodynamic simulations (usually with 
{\it temperature--independent} $\eta/s$) have shown that {\em large enough} means $\eta/s \sim 0.3-0.5$ 
\cite{apli,shen,tor,REVS,bulk1,bulk2,bulk3,song}. For larger values of $\eta/s$ it is not possible to match the 
predictions of hydrodynamic models to RHIC data. Moreover, it has been shown by direct 
comparison to results obtained with Boltzmann equation that for matter created at RHIC, viscous 
hydrodynamics fails to reproduce the results of Boltzmann equation if $\eta/s \gtrsim 0.2$ 
\cite{MOLNAR}. Beyond this value, an appropriate treatment of the dynamics of this matter 
should switch smoothly from a viscous hydrodynamic description to a kinetic one, in which the 
issue of breakdown due to large viscosity (i.e. dissipation) does not arise. For recent 
developments and applications of this hybrid description see Refs.~\cite{song,shen}. 

Since the model used here is purely hydrodynamic, a cut--off for the value of $\eta/s$ must be 
{\it imposed} in order to avoid the breakdown of the fluid description. 
The value of $\eta/s$ at which it is sensible to impose this cut--off is 
constrained by the comparison of results obtained from hydrodynamic and kinetic simulations to 
data. For this reason, in our simulations we set $\eta/s \leq 0.4$, which corresponds to $T 
\sim 140$ MeV for the three values of the chiral coupling, namely
$g=3,3.2,3.4$, the values we consider in the present paper. 
We emphasize that in all of 
our simulations with $g=3,3.2,3.4$ the {\it average} value of $\eta/s$ throughout the 
hydrodynamic evolution is $\sim 0.11$, and thus it remains well below the value at which 
viscous hydrodynamics breaks down ($\eta/s \sim 0.3-0.5$).

In Appendix~\ref{cutoff} we analyse the dependence of our results on the choice of different 
cut--off values for $\eta/s$. We find that, although the values of $v_2$ and $v_4/(v_2)^2$ do 
change, the differences between the cases including or not the chiral fields as sources remain  
essentially the same, and thus our main conclusions are not sensitive to the value imposed for 
the cut--off.

\begin{figure}[htb]
\scalebox{0.65}{\includegraphics{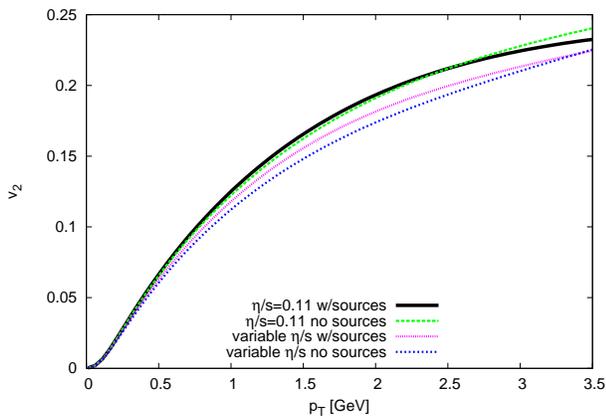}}
\caption{(Color online) Elliptic flow of charged--hadrons calculated taking or not taking into 
account the source terms in the hydrodynamic equations, for either a temperature--dependent or 
a temperature--independent $\eta/s$ which is the averaged value throughout the evolution of the fireball. The 
value of the coupling constant is $g=3.2$.}
\label{v2}
\end{figure}

Knowing $c_s^2$ and $\eta/s$ as functions of the temperature, we can now go over to discuss the momentum 
anisotropies obtained from the simulations. Fig.~\ref{v2} shows the charged--hadron elliptic flow $v_2$ 
calculated with either the temperature--dependent $\eta/s$ of the LSM or a temperature--independent 
$\eta/s = 0.11$, which corresponds to the average of $\eta/s$ throughout the hydrodynamic evolution over
the temperature interval from $T_i$ to $T_f$. In both cases, we set $g=3.2$ and compare the results 
obtained by taking or not taking into account the source terms in the hydrodynamic equations. 

It is seen from Fig. \ref{v2} that both for a temperature--dependent and a temperature--independent $\eta/s$, 
the elliptic flow does not depend strongly on the chiral sources. For a temperature--independent 
$\eta/s$, $v_2$ turns out to be practically independent of the dynamics of the chiral sources for 
$p_T~<~3$~GeV. In contrast, 
for the temperature--dependent $\eta/s$ there are small, but visible differences in the elliptic flow 
calculated with or without the chiral sources. Specifically, $v_2$ is slightly larger and reaches a maximum 
at $p_T \sim 2$~GeV when the chiral fields are taken into account as sources for the hydrodynamic
variables. 

Comparing the behavior of $v_2$ for temperature--independent or temperature--dependent $\eta/s$, it is seen that a 
temperature--dependent $\eta/s$ leads to a smaller elliptic flow. This fact has 
already been seen in the studies of Refs.~\cite{niemi,bozek,{Nagle:2011uz}} by performing
hydrodynamic simulations (not coupled to chiral fields) with different parameterizations 
for temperature--dependent $\eta/s$ (usually a step--function dependence or some slight
variation), and in Ref.~\cite{song} from a Boltzmann--hydrodynamics hybrid approach. See also 
Refs.~\cite{koide,shen} for related studies. In particular, it was found in Ref.~\cite{niemi}
that $v_2$ at Au$+$Au collisions at the largest RHIC energy is dominated by the shear 
viscosity in the hadronic phase and largely insensitive to the viscosity of the QGP. 
A similar conclusion was reached in Refs.~\cite{bozek,{Nagle:2011uz}}. 
It is interesting to note that, according to recent viscous hydrodynamic simulations with temperature 
dependent $\eta/s$, at LHC energies the situation is exactly the opposite, i.e. the elliptic flow 
becomes sensitive to the QGP viscosity and insensitive to the hadronic viscosity~\cite{Nagle:2011uz,niemi}.
As noted in Ref.~\cite{Nagle:2011uz}, the fact $v_2$ depends rather weakly on the temperature dependence 
of $\eta/s$ poses serious challenges to the precise extraction of this ratio from this 
hadronic observable.
For this reason, and also due to the uncertainty in the initial conditions in 
heavy--ion collisions at RHIC and LHC, it has now become important to calculate higher--order  
Fourier moments $v_3$, $\cdots$ in viscous hydrodynamic simulations, which may provide further crucial 
constraints on $\eta/s$ and on models used to calculate QGP initial conditions -- see e.g. 
Refs.~\cite{song,higher1,higher2,higher3,higher4,higher5,higher6}.

Returning to our results in Fig.~\ref{v2}, it seems clear the difficulties in extracting (an average 
value of) $\eta/s$ from data on $v_2$. The results show that uncertainties associated with the dependence 
of $\eta/s$ on temperature lead to appreciable changes in the curve of $v_2$ versus $p_T$. Such
an uncertainty should be added to the theoretical uncertainty that comes e.g. from the 
initial conditions (for example using Color Glass Condensate or Glauber initial conditions), 
the freeze--out process (as issues concerning Grad's quadratic ansatz), as well as from other 
sources, that according to recent studies add up to an overall uncertainty which can be roughly 
estimated in $0.1$~\cite{luzum,PRC}. 

As mentioned in the Introduction, the fact that $v_2$ depends somewhat on the evolution of 
the chiral fields in the case of a temperature--dependent $\eta/s$ has already been 
seen in the study by Plumari et al \cite{Plumari:2010ah} within Boltzmann--Vlasov simulations on 
the NJL model. Our results show that this also happens in a purely hydrodynamic simulation as
well. 

\begin{figure}[htb]
\scalebox{0.65}{\includegraphics{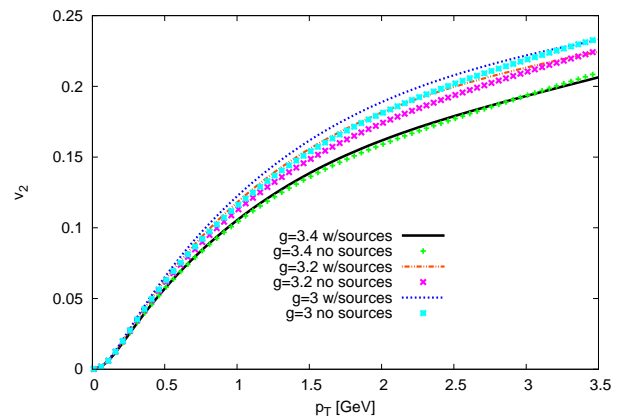}}
\caption{(Color online) Elliptic flow of charged--hadrons calculated taking or not 
taking into account the source terms in the hydrodynamic equations, for $g=3,3.2,3.4$ 
and a temperature--dependent $\eta/s$.}
\label{v2g}
\end{figure}

Since our approach is phenomenological, it is important to determine the sensitivity of our 
conclusions regarding the influence of the chiral fields on $v_2$ to the value of the coupling 
constant $g$. Fig. \ref{v2g} shows the elliptic flow of charged--hadrons calculated taking or 
not taking into account the source terms in the hydrodynamic equations, for $g=3,3.2,3.4$ and a 
temperature--dependent $\eta/s$. It is seen that when the value of $g$ increases, $v_2$ 
decreases. 
This is due to the fact that for larger values of $g$, $c_s^2$ near the transition region becomes smaller 
which means that pressure gradients are converted into flow less efficiently, thus resulting in lower values 
of $v_2$.  
The difference in the values of $v_2$ calculated including or not the chiral fields as sources is 
seen to become smaller with increasing $g$, which supports the conclusion reached earlier that 
the influence of the chiral fields on $v_2$ is rather small.

\begin{figure}[htb]
\scalebox{0.675}{\includegraphics{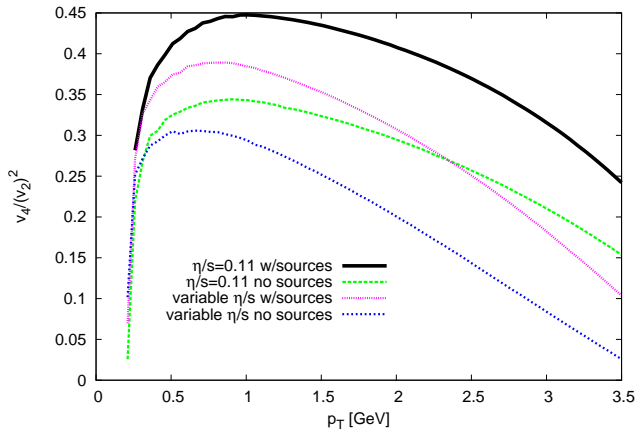}}
\caption{(Color online) $v_4/(v_2)^2$ for charged--hadrons calculated taking or not taking into 
account the chiral fields as sources for the hydrodynamic equations, for either a temperature--
dependent or a temperature--independent $\eta/s$. The value of the coupling constant is $g=3.2$.}
\label{v4}
\end{figure}

\begin{figure}[htb]
\scalebox{0.675}{\includegraphics{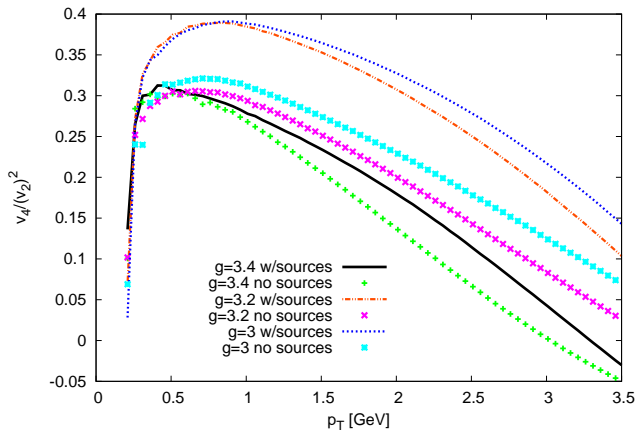}}
\caption{(Color online) $v_4/(v_2)^2$ for charged--hadrons calculated taking or not taking into 
account the chiral fields as sources for the hydrodynamic equations, for $g=3,3.2,3.4$ and a 
temperature--dependent $\eta/s$.}
\label{v4g}
\end{figure}

Another interesting observable to analyse is $v_4/(v_2)^2$, which has only recently been 
theoretically investigated. See in particular Refs.~\cite{v4luzum1,v4luzum2} for a 
detailed account of the physics (and caveats -- see below) involved in this observable within dissipative hydrodynamics. 
See also Ref.~\cite{Plumari:2010ah} for a calculation of $v_4/(v_2)^2$ within Boltzmann--Vlasov 
formalism. Fig.~\ref{v4} shows $v_4/(v_2)^2$ for the same cases as in Fig. \ref{v2}. 
In contrast to what happens with $v_2$, it is seen that $v_4/(v_2)^2$ is significantly 
affected by the dynamics of the chiral fields even at low transverse momentum and for a temperature--independent $\eta/s$. 
In particular, the values of $v_4/(v_2)^2$ are larger when the chiral fields are included as sources in the
hydrodynamic equations, and the behavior with $p_T$ is somewhat different too. It is also seen 
that a temperature--independent $\eta/s$ leads to larger $v_4/(v_2)^2$.

At this point, a comment on the possible extraction of $\eta/s$ from $v_4/(v_2)^2$ data by
matching to hydrodynamic simulations is in order. It is known that the dependence of 
$v_4/(v_2)^2$ on tranvserse momentum that is obtained from viscous hydrodynamics does 
not agree with the almost constant value of $v_4/(v_2)^2 \sim 1$ measured at RHIC (see Refs.~\cite{v4luzum1,v4luzum2}). 
Ideal hydrodynamics predicts $v_4/(v_2)^2=0.5$, but quite 
surprisingly viscous hydrodynamics yields a strongly dependent ratio, as shown, for
example, in Fig.~\ref{v4}. A possible explanation for this discrepancy has been put 
forward by Luzum and Ollitrault~\cite{v4luzum1,v4luzum2}, who realized that Grad's quadratic ansatz 
for the nonequilibrium correction $\delta f$ to the particle distribution fuction may not 
be valid for the freeze--out process in heavy--ion collisions. See also 
Refs.~\cite{gradansatz1,gradansatz2,gradansatz3,gradansatz4} for recent work on the reliability of Grad's ansatz. 
By performing viscous hydrodynamic simulations with different dependencies of $\delta f$ on $p_T$, the 
authors of Refs.~\cite{v4luzum1,v4luzum2} have 
found that RHIC data on $v_4/(v_2)^2$ favors a momentum dependence between linear and 
quadratic. In spite of this, we have chosen to show results on $v_4/(v_2)^2$ to emphasize 
that, although the values for this observable are very different from those measured at 
RHIC (because we are using Grad's quadratic ansatz at freeze--out), they are much more 
sensitive than $v_2$ to the evolution of the chiral fields. We expect this feature of 
$v_4/(v_2)^2$ to hold when other forms for $\delta f$ are used in the simulations, 
although further work is needed to confirm this expectation. 

As with $v_2$, it is important to determine the dependence of our results and conclusions 
on the value of $g$. Fig.~\ref{v4g} shows $v_4/(v_2)^2$ calculated including or not the 
sources in the hydrodynamic equations, for $g=3,3.2,3.4$ and a temperature--dependent 
$\eta/s$. It is seen that the difference in the values of $v_4/(v_2)^2$ calculated including 
or not the chiral fields as sources becomes smaller with increasing $g$, but is still 
relatively large at $g=3.4$. Moreover, it is seen that the behavior of this ratio with 
$p_T$ does not depend strongly on the value of $g$. These results obtained with different 
values of $g$ support the conclusion mentioned before that $v_4/(v_2)^2$ is more sensitive 
than $v_2$ to the influence of the chiral fields on the evolution of the quark fluid.

\section{Conclusions}
\label{conc}

We have studied the evolution of the fireball created at RHIC within second--order 
viscous hydrodynamics coupled self--consistently to the LSM, focusing on 
the impact of the dynamics of the chiral fields on the hadronic observables $v_2$ and $v_4/
(v_2)^2$. We have compared the results obtained when the chiral fields are included or not as 
sources in the hydrodynamic equations. The comparisons were made using both 
temperature--independent and temperature--dependent $\eta/s$. The temperature dependence 
of $\eta/s$ was calculated in the LSM using the linearized Boltzmann equation.  

We have found that the values of $v_2$ do not depend strongly on the evolution of the 
chiral fields. Specifically, for a temperature--independent $\eta/s$ this dependence 
is negligible for $p_T < 3$ GeV, while for a temperature--dependent $\eta/s$ it is 
appreciable but still small even at small $p_T$. We have also found that the ratio 
$v_4/(v_2)^2$ is much more sensitive to the dynamics of the chiral fields, being larger 
when these fields are taken into account as sources in the hydrodynamic equations, 
in both situations of the temperature dependence of $\eta/s$. 

In line with the results of Refs.~\cite{Nagle:2011uz,niemi,shen}, 
our results show that not knowing precisely the temperature--dependence of $\eta/s$ leads 
to further uncertainties in attempts of extracting this ratio from data on $v_2$, in
addition to the uncertainties that stem from the initial conditions and the
freeze--out process, among others sources.  It is worth noting that despite the 
coupling of chiral sources to the hydrodynamic evolution would add further uncertainties, 
they are not very big.  

The model used in this work leaves room for improvements in different directions. Probably, the most 
important ones for our analysis are including bulk viscosity in the hydrodynamic equations and fluctuations 
of the chiral fields. This latter effect would act as noise sources in the classical equations of motion. We believe 
that the qualitative trends and the general conclusions extracted from our results will hold when these 
effects are taken into account. Work is in progress where these two aspects are taken
into account in a viscous hydrodynamic simulation and their impact on observables will 
be reported in a forthcoming publication. 

\appendix

\section{Linear $\sigma$ model}
\label{lsms}

In this Appendix we will briefly review the relevant aspects of linear $\sigma$ model for the
present paper, as well as the calculation of $\eta$ from the linearized Boltzmann equation in the relaxation 
time approximation. 

As mentioned before, as an effective theory of the chiral symmetry breaking dynamics we 
consider the linear $\sigma$ model coupled to two flavors of constitutive quarks. The Lagrangian density of the coupled system is given 
in Eq.~(\ref{lag}). 
The quark fluid is considered as a thermal bath for the chiral field, and therefore it can be integrated out to obtain the effective 
potential $V_e$ for the
chiral fields in the presence of that bath of quarks. 
In order to calculate the equilibrium pressure density $p(\phi,T)=-V_e(\phi,T)+U(\phi)$, we use the one-loop effective 
potential~\cite{paech1,paech2}:
\begin{equation}
V_e (\phi,T) = U(\phi) - d_q T \int \frac{d^3p}{(2\pi)^3}
\, \ln \left(1+e^{-E/T}\right) ~,
\end{equation}
where $d_q = 24$ is the color--spin--isospin--baryon charge degeneracy of the quarks and 
the energy $E=(p^2+m_q^2)^{1/2}$ with $m_q^2 = g^2\phi^2$. The equilibrium energy density is just 
\begin{equation}
\epsilon(\phi,T)= d_q \int \frac{d^3 p}{(2\pi)^3} f_0 ~E ~.
\end{equation}
From $p$ and $\epsilon$, the square of the speed of sound and the entropy density follow: $c_s^2 = d p/d \epsilon$ and $s=\partial p/\partial T$.

%
%
The local 
thermal averages of the scalar and pseudo-vector chiral densities 
$\rho_s = \langle \bar q q \rangle$ and $\vec \rho_{ps} = \langle \bar q \gamma_5 \vec \tau q \rangle$ enter as sources in the hydrodynamic equations. 
Explicitly, they are given by:
%
%
\bea
\rho_s &=& g \sigma d_q \int \frac{d^3 p}{(2\pi)^3} \frac{1}{E} f_0 ,\\
\vec{\rho}_{ps} &=& g \vec{\pi} d_q \int \frac{d^3 p}{(2\pi)^3} \frac{1}{E} f_0 ~.
\label{rhosrhops}
\eea
%

The LSM exhibits a first--order phase transition, a crossover and a critical end point, depending on the value of 
the chiral coupling constant $g$. The crossover phase transition is not a genuine phase transition since all
thermodynamic functions change smoothly with temperature. However, such changes may be quite sudden in a narrow temperature range. 
As indicated in Section~\ref{res}, throughout this work we set $g=3$, $g=3.2$ and $g=3.4$, which correspond to a smooth crossover 
and lead to a temperature--dependent sound speed that resembles that of Lattice QCD. 
\begin{figure}[htb]
\scalebox{0.675}{\includegraphics{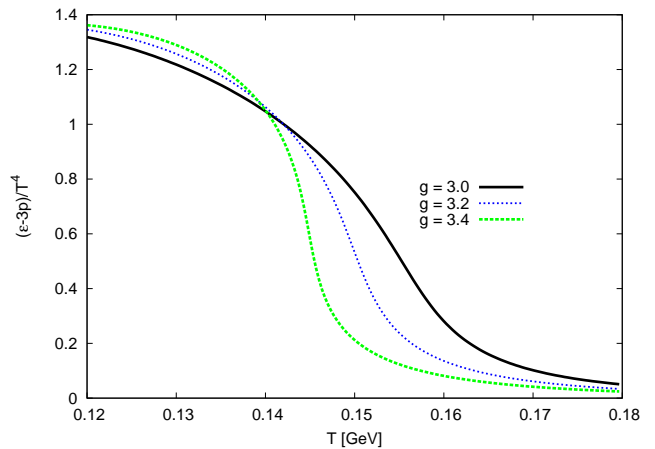}}
\caption{(Color online) $(\epsilon - 3p)/T^4$ of the LSM as a function of temperature for $g=3,3.2,3.4$.}
\label{pe}
\end{figure}

Here we model the expansion of the QGP using viscous hydrodynamics with the  
bulk viscosity $\zeta$ set to zero. It is known 
that $\zeta$ scales as $c_s^2-1/3$, so from Fig. \ref{cs}  it is clear that for the LSM 
$\zeta \neq 0$ (see also Refs.~\cite{trans-kinetic5,kap}). Recent Lattice QCD results indicate that the
quark-gluon matter EOS departures from the conformal description \cite{tc1,tc2,tc3,tc4,tc5}, 
although it is common practice to neglect $\zeta$ in the hydrodynamic equations while 
using an EOS inspired on Lattice QCD. 
An estimate for the temperature--dependence and the relative importance of $\zeta$ can be obtained from $(\epsilon-3p)/T^4$, 
which measures the deviation from the conformal limit. Fig. \ref{pe} shows this quantity calculated in the LSM as a function 
of temperature, for $g=3,3.2,3.4$. 
It is seen that, for the range of temperatures of interest $T\geq 130$ MeV, $(\epsilon-3p)/T^4$ is small except at $T\leq T_c$ where it rises 
abruptely reaching significant values. The increase in $(\epsilon-3p)/T^4$ near $T_c$ is sharper for larger values of $g$. It has been shown 
before that the hydrodynamic evolution is affected by bulk viscosity, for example reducing the elliptic flow, inducing the phenomenon of 
cavitation or modifying the freeze--out process \cite{bulk1,bulk2,bulk3}. Although it is reasonable to assume that these changes will not affect 
in a dramatic way the differences in $v_2$ and $v_4/(v_2)^2$ that result from including or not the chiral fields as sources for the evolution of 
the quark fluid, a more thorough investigation of this particular issue is necessary. 

We now go over to the calculation of the shear viscosity. In order to calculate $\eta$ we adopt the results 
of Ref~\cite{trans-kinetic5}, which are based on the linearized Boltzmann equation in the relaxation time 
approximation. In the relaxation time approximation the shear viscosity is 
given by
\begin{equation}
\eta = \frac{4\tau}{5T}\int  \frac{d^3p}{(2\pi)^3} \frac{p^4}{E^2}f_0 (1-f_0)
\end{equation}
where $\tau=\tau(T)$ is the collision time.  $\tau(T)$ is calculated from the averaged cross sections $\bar{\sigma}$ 
for quark--quark and quark--antiquark scattering processes including $1/N_c$ next to leading order corrections as:
\bea
\tau^{-1} &=& 6 f_0 \Bigl( \bar{\sigma}_{uu\rightarrow uu} + \bar{\sigma}_{ud\rightarrow ud}
+ \bar{\sigma}_{u\bar{u}\rightarrow u\bar{u}} + \bar{\sigma}_{u\bar{u}\rightarrow d\bar{d}}
\nn\\
&& + \, \bar{\sigma}_{u\bar{d}\rightarrow u\bar{d}}\Bigr) ~~.
\eea
We refer the reader to Refs.~\cite{trans-kinetic5,sigmas1,sigmas2} for details on the
calculation of the $\bar{\sigma}'s$ -- we note that the chiral model used in these references 
is not very different from the LSM, the exchanges of $\sigma$ and $\pi$ mesons are modeled 
by contact terms. Note also that the cross sections are temperature dependent, not only 
because of phase space, but also because they depend on the constituent quark masses,
whose temperature dependence is given by the LSM of the present paper. 
See also Ref.~\cite{kap} for a related approach applied to the calculation of $\eta/s$ in 
the LSM of a pion gas.

\section{Dependence of results on cut--off for $\eta/s$}
\label{cutoff}

It was noted before that any hydrodynamical description of matter is bound to become
inapplicable at large viscosity.  Since our simulations are purely hydrodynamic and the
calculated $\eta/s$ increases rapidly with decreasing temperature, a cut--off for the value 
of $\eta/s$ must be imposed. The comparison of the results obtained from hydrodynamic and
kinetic simulations to data provide a guide as to what is the value at which one should impose
the cut--off. In our simulations we set the cut--off at $\eta/s = 0.4$, which is in the range of
values found to correspond to the breakdown of viscous hydrodynamics \cite{REVS,apli}. In this
Appendix we show the results for hadronic observables calculated using different values of this
cut--off, namely $\eta/s \leq 0.3,0.4$ and $0.5$, and analyse the  dependence of the results 
on this choice for the cut--off. For simplicity, the results shown in this appendix correspond
to $g=3.2$.

\begin{figure}[htb]
\scalebox{0.675}{\includegraphics{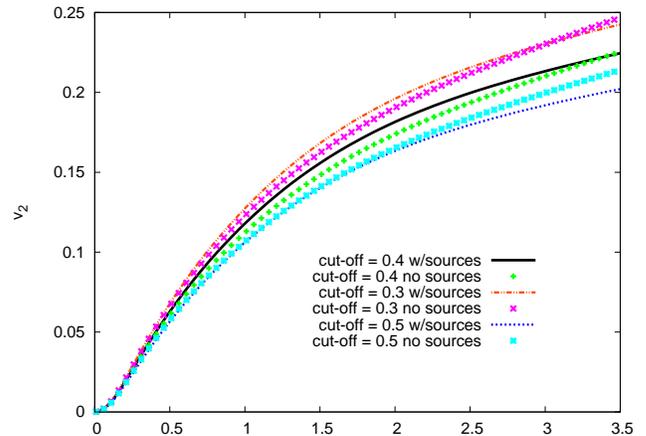}}
\caption{(Color online) $v_2$ for charged--hadrons calculated taking or not taking into account 
the chiral fields as sources for the hydrodynamic equations calculated with different choices 
for the cut--off value $\eta/s=0.3,0.4,0.5$. The value of the coupling constant is $g=3.2$.}
\label{v2c}
\end{figure}

\begin{figure}[htb]
\scalebox{0.675}{\includegraphics{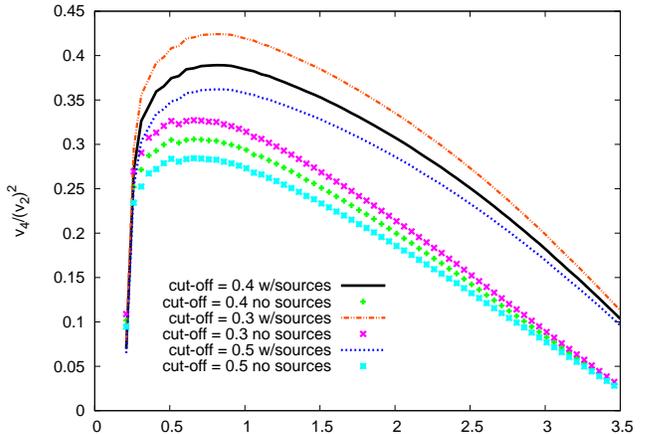}}
\caption{(Color online) $v_4/(v_2)^2$ for charged--hadrons calculated taking or not taking into 
account the chiral fields as sources for the hydrodynamic equations calculated with different 
choices for the cut--off value $\eta/s=0.3,0.4,0.5$. The value of the coupling constant is 
$g=3.2$.}
\label{v4c}
\end{figure}

Figs. \ref{v2c} and \ref{v4c} show the charged--hadron $v_2$ and $v_4/(v_2)^2$, respectively, 
as a function of transverse momentum, obtained with cut--off values of $\eta/s = 0.3,0.4$ and 
$0.5$. It is seen that although the values of $v_2$ and $v_4/(v_2)^2$ change with the value of 
the cut--off, the relation between the results obtained including or not the chiral fields as 
sources in the hydrodynamic equations remains practically the same. Consequently, the 
conclusions extracted from these results, which are discussed in the main text, do not depend on 
the precise value of the cut--off imposed on $\eta/s$, provided that $g\sim 3.2$ corresponding to a 
smooth crossover. 

As a final side remark, it is interesting to note that if one would attempt to 
extract the value of $\eta/s$ by matching $v_2$ to data (with the remark mentioned in 
the Introduction), the uncertainty in the extracted $\eta/s$ coming from the choice of 
cut--off in the range $[0.3,0.5]$ would be $\sim 50 \%$ with respect to the average value of $\eta/s$. This relatively large value of the
uncertainty in $\eta/s$ associated with the imposed cut--off clearly displays the sensitivity 
of observables on the shear viscosity of the hadronic stage in collisions at $\sqrt{s_{NN}} = 200$~GeV,
a fact already discussed in Refs.~\cite{niemi,bozek,{Nagle:2011uz},shen}, pointing to the conclusion that further study 
is needed to extract the precise temperature dependence of $\eta/s$ from heavy--ion collisions experiments. 

\begin{acknowledgments}
We thank Jorge Noronha for useful comments on the calculation of transport coefficients of stronly coupled theories. This work was partially funded by CNPq and FAPESP (Brazilian agencies). 
\end{acknowledgments}

\end{document}